\documentclass{article}

\usepackage[preprint]{neurips_2023}

\usepackage[utf8]{inputenc} 
\usepackage[T1]{fontenc}    
\usepackage{hyperref}       
\usepackage{url}            
\usepackage{booktabs}       
\usepackage{amsfonts}       
\usepackage{nicefrac}       
\usepackage{microtype}      
\usepackage{xcolor}         
\usepackage{graphicx}

\title{Dynamic Grouping for Climate Change Negotiation: Facilitating Cooperation and Balancing Interests through Effective Strategies}

\author{Yu Qin \\
Eccles School of Business \\
University of Utah \\
Salt Lake City, UT, U.S.A. \\
\texttt{yu.qin@eccles.utah.edu} 
\And 
Duo Zhang \\
McGill University \\
845 Rue Sherbrooke O, Montréal, QC H3A 0G4, Canada \\
\texttt{duo.zhang4@mail.mcgill.ca}
\And
Yuren Pang \\
University of Washington \\
Seattle, WA, USA \\
\texttt{ypang2@uw.edu}
}

\begin{document}

\maketitle

\begin{abstract}
In this paper, we propose a dynamic grouping negotiation model for climate mitigation based on real-world business and political negotiation protocols. Within the AI4GCC competition framework, we develop a three-stage process: group formation and updates, intra-group negotiation, and inter-group negotiation. Our model promotes efficient and effective cooperation between various stakeholders to achieve global climate change objectives. By implementing a group-forming method and group updating strategy, we address the complexities and imbalances in multi-region climate negotiations. Intra-group negotiations ensure that all members contribute to mitigation efforts, while inter-group negotiations use the proposal-evaluation framework to set mitigation and savings rates. We demonstrate our negotiation model within the RICE-N framework, illustrating a promising approach for facilitating international cooperation on climate change mitigation.
\end{abstract}

\section{Introduction and Theoretical Background}

Climate change is an increasingly pressing issue with far-reaching consequences for both human and natural systems, causing increasing global temperature, rising sea levels, more extreme weather, and damaging human livelihood. 
For example, the devastating wildfires that ravaged Australia in 2019-2020, fueled by record-breaking temperatures and prolonged drought, destroyed millions of hectares of land and killed billions of animals~\cite{knuthwebsite}. In the year 2022, Pakistan was struck by a severe flood event, leading to extensive inundation covering a significant portion of the country~\cite{philanthropy_2023}. The catastrophic event caused over 1500 fatalities, while rendering a staggering 33 million people homeless. During the summer of 2022, the Yangtze River Basin and its surrounding areas in China were struck by prolonged and intense heatwaves, resulting in several secondary disasters such as droughts, wildfires, and power shortages~\cite{scmpChinasRecordbreaking}.
Recent alarming extreme weather headlines around the world reminded us of the urgent need to address climate change. Failure to take action will result in irreversible damage to the planet and its inhabitants. We need to take action.

Tackling climate change necessitates \textit{global cooperation}, as climate actions by individual countries would likely have an impact beyond the country border. To facilitate such international collaboration, the United Nations Framework Convention on Climate Change (UNFCCC) establishes an international environmental treaty~\cite{UNFCCC1992}, which has been signed by 165 countries. The Paris Agreement, adopted in 2015 under the UNFCCC~\cite{ParisAgreement2015}, sets a goal of limiting global warming to well below 2 degrees Celsius above pre-industrial levels and pursuing efforts to limit it to 1.5 degrees Celsius. The agreement also requires countries to submit nationally determined contributions (NDCs) outlining their targets and actions to reduce emissions and adapt to climate change. However, achieving the goals set in the Paris Agreement will require not only ambitious targets and actions from individual countries but also collective efforts to support developing countries in their transition to low-carbon and climate-resilient pathways~\cite{zhang2022ai}. Furthermore, effective implementation of climate policies and measures will require cooperation among multiple stakeholders, including governments, private sectors, the public, and international organizations. 

Under the backdrop of global momentum, \textit{regional cooperation} should also play an important role to facilitate climate negotiation. As there is a significant imbalance in resources and capacity among countries, regional organizations---such as the ~\cite{asean_declaration}, ~\cite{AU2000}, and the ~\cite{UNASUR2008}---have been formed to protect regional interests. Such regional collaboration is especially important for smaller and developing countries, as they often lack the financial and technological resources to implement ambitious climate policies and measures on their own~\cite{vinogradov2020adaptation}. %
As such, such real-world collaboration among these countries is critical to advance climate action at the global level. For example, the African Union has established a Climate Change Fund to support adaptation and mitigation activities in African countries~\cite{ACCF}. The Pacific Islands Forum has called for urgent action to limit global warming to 1.5 degrees Celsius, given the vulnerability of Pacific Island countries to climate impacts~\cite{PacificIslandLeaders2021}.

In our proposal, we propose to account for regional group collaboration in a \textit{dynamic grouping model} in the global climate change framework using the existing \texttt{RICE-N}~\cite{zhang2022ai}. The assumption is that climate negotiation can be swayed in a negative by powerful countries in the original framework. Concretely, we grouped similar countries into regional units, augmenting the voice of smaller countries in the RL framework. By working together, smaller countries can amplify their voices and leverage their resources to push for more ambitious climate action at the international level and ensure that the global response to climate change is equitable and inclusive. Our evaluation suggests that the proposed dynamic grouping model can reduce global temperature rise, as well as increasing economic growth for each region. Our proposal serves as a practical yet effective measure to improve global climate change negotiation.


\section{Methodology Development}

Inspired by real-world business and political negotiation protocols~\cite{goodman2002politics, dash2008regionalism}, we have developed, tested, and implemented a dynamic grouping negotiation model for climate mitigation within the AI4GCC competition framework. The model aims to facilitate efficient and effective cooperation between various stakeholders to achieve global climate change objectives. This section will detail the key components of the model, such as group formation and updates, intra-group negotiation, and inter-group negotiation. Additionally, we will present the end-to-end negotiation workflow within the context of the \texttt{RICE-N} framework. The framework is visualized in Figure 1.

\begin{figure}[h]
    \centering
    \includegraphics[width=0.8\textwidth]{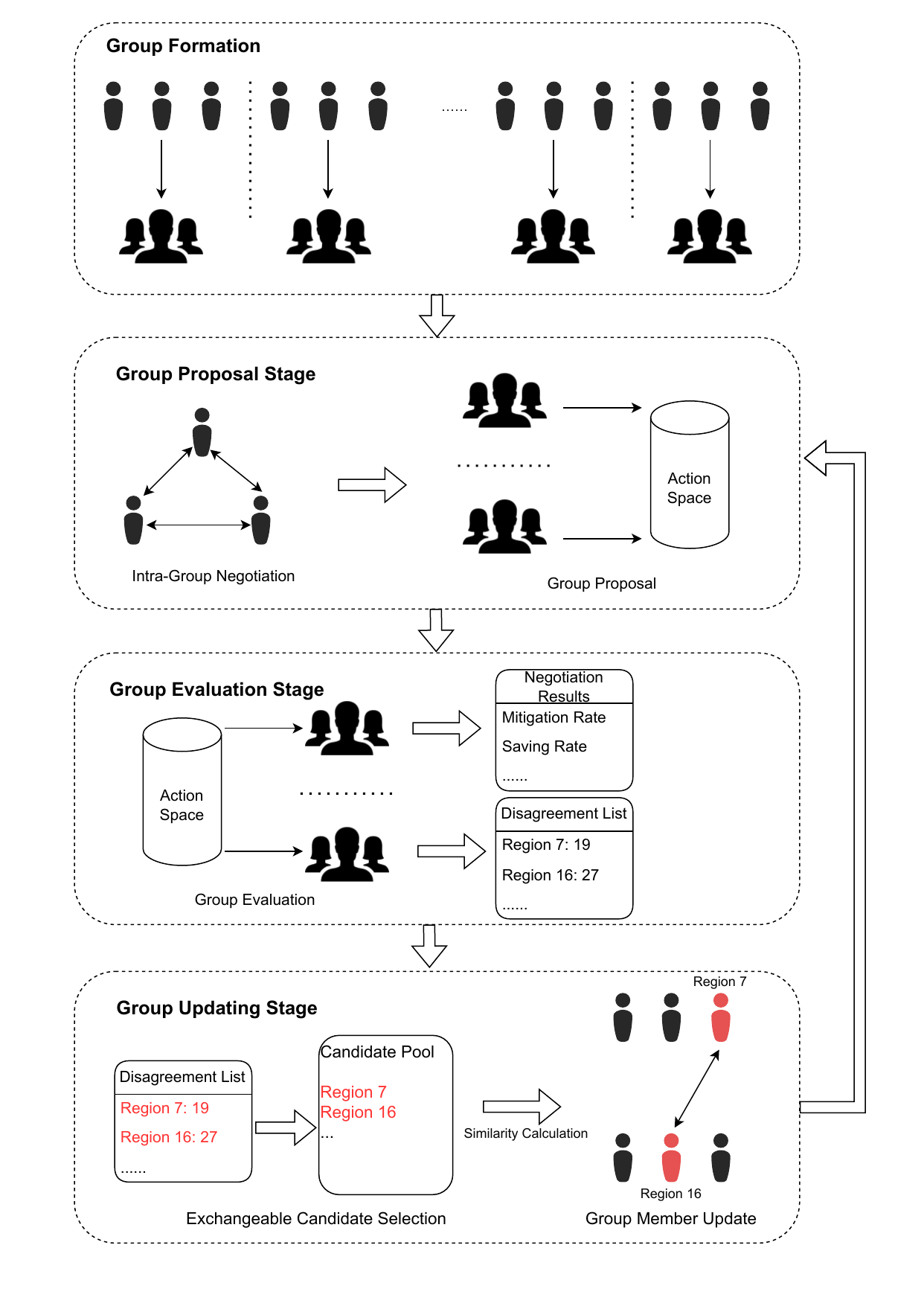}
    \caption{Dynamic Grouping Framework}
    \label{fig:label_of_the_figure}
\end{figure}

\subsection{Group Formation and Updates}

Rather than using the bilateral negotiation setting from the baseline, we propose that a group setting is more appropriate for multi-region climate negotiations. First, each region has varying levels of population and capital. The unbalanced resources make it difficult for smaller or developing regions to be treated equally during bilateral negotiations. In reality, small countries often participate in international activities as part of a group, allowing them to better protect their interests and foster a more stable collaborative environment. Consequently, we introduce a group-forming method that helps the 27 regions in this simulated environment form three-region groups, resulting in nine regional groups. Initial group membership is assigned based on each region's population and capital, using the following principles:

1. Assign one region with adequate capital to each group to support the group's economic growth.\\
2. Ensure a similar population distribution across different groups. Specifically, we first guarantee that there is at least one region with a large population to provide the necessary human resources for group development. Then, we assign regions with smaller populations to each group to achieve even cross-region population distribution, taking capital into account.

Following these principles, we form nine groups within the 27 regions. The initialized groups are shown in Table 1.

\begin{table}[ht]
\centering
\begin{tabular}{@{}ccccc@{}}
\toprule
\textbf{Group} & \multicolumn{4}{c}{\textbf{Region IDs}} \\ \midrule
0              & 26          & 1            & 2            &             \\
1              & 3           & 4            & 6            &             \\
2              & 5           & 7            & 18           &             \\
3              & 19          & 10           & 12           &             \\
4              & 20          & 13           & 15           &             \\
5              & 14          & 16           & 22           &             \\
6              & 8           & 9            & 21           &             \\
7              & 11          & 17           & 23           &             \\
8              & 24          & 25           & 0            &             \\ \bottomrule
\end{tabular}
\caption{Initialized regional groups}
\label{table:regional_unions}
\end{table}

With our initial group assignments, the negotiation process can begin. However, it is nearly impossible for group members to remain static over a 50-year horizon. If collaboration doesn't work well, it's natural for group members to be updated or new groups to form to achieve better intra-group collaboration. As a result, we have designed a group member updating strategy that allows each group to change members if needed. Specifically, we maintain a list to track the decisions made by each group member during the negotiation. If a group member makes different decisions than the other two regions in the group for a certain time, we place it in the exchangeable candidate pool, awaiting potential group switching. If there are two regions in the exchangeable candidate pool with similar population and capital, we exchange their groups to see if the updated group leads to better outcomes. We define similarity in population and capital by calculating the sum of the absolute second-order differences across two indicators. If the total difference is smaller than a certain threshold, the regions are classified as similar.

\subsection{Intra-group negotiation}

As mentioned earlier, each group consists of three regions with varying resources and development levels, resulting in different abilities to afford mitigation rates. While the primary negotiation occurs between groups, it is crucial for members within a group to discuss their shares of the mitigation responsibility. In the bilateral scenario, each region has a choice of mitigation levels from 0 to 10 before normalization. Following this setting, we assign each group a mitigation choice ranging from 0 to 10. However, this mitigation level indicates the group-average mitigation, meaning that we ask the regions within the group to achieve an average mitigation at this level. Consequently, members of each group must discuss the mitigation they will undertake to contribute to the group's overall mitigation. We do not include intensity in this calculation for the following reasons:

1.  Separating mitigation based on intensity may encourage smaller countries to avoid mitigation and transfer responsibility to larger countries, which is not the desired outcome.\\
2.  Since the final intensity will ultimately influence climate change and, consequently, the results, large countries cannot transfer their mitigation to smaller countries, as this would cause a significant temperature increase in the future.

Under our approach, all three regions in each group will strive to contribute to mitigation, thereby contributing to global climate change.

\subsection{Inter-group negotiation}

We employ the proposal-evaluation framework from bilateral negotiations to facilitate inter-group negotiations. Instead of 27 regions, we now have 9 groups, with negotiations occurring among these groups. First, each group proposes a set of requirements for mitigation and savings rates for all groups. It is worth noting that, in addition to the mitigation rate from the baseline model, we have incorporated a savings rate into the negotiation process. We believe that, within a global collaboration framework, it is insufficient to only control mitigation for climate change. Appropriate economic constraints can facilitate negotiations and provide further motivation for mitigation as potential economic sanctions.

Once each group has proposed its requirements, they exchange information and evaluate the proposals. During the evaluation stage, each region decides whether to accept the offer from other groups or not. Importantly, a region cannot reject the mitigation requirement while accepting the savings rate requirement, or vice versa. These two requirements must be accepted or rejected simultaneously to ensure that the savings rate influences a region's judgment on mitigation. After each region has made its decision, an intra-group voting process takes place. If at least two regions in a group accept the requirements, they are accepted by the group. Conversely, if only one region accepts the requirements, they are rejected. We then set the minimum mitigation rate and savings rate based on the evaluation results, following the method used in bilateral setting.

\subsection{Negotiation Workflow}

We have developed a negotiation workflow within the RICE framework based on the negotiation protocol described earlier. Overall, the model has three stages in a complete iteration: group proposal stage, group evaluation stage, and group updating stage.

\textbf{Group proposal stage}: In this stage, the intra-group negotiation takes place first. Each group member discusses their share of the mitigation responsibility. Once the shares are settled, they remain fixed until the next iteration begins. Subsequently, each region proposes its requirements for mitigation and savings rates as part of the inter-group negotiation content. These proposals are recorded and await further evaluation.

\textbf{Group evaluation stage}: In this stage, the inter-group negotiation continues. As previously discussed, each region evaluates the proposals from other groups and makes its own decision to accept or reject the proposal. Then, an intra-group vote determines the final evaluation results and sets the minimum average mitigation and savings rates for each group. The decision results are also recorded for future group updating purposes. Based on the evaluation results, the three regions in each group divide the shares according to the intra-group negotiation results, ensuring that their average mitigation and savings rates meet the mutually agreed-upon outcome.

\textbf{Group updating stage}: In this stage, we examine the group decision results and update groups if certain conditions are met. First, we review the decision results in each group and add the number of disagreements to a list that tracks each region's inconsistencies within the group. If the number of inconsistencies exceeds a specified threshold (18 in our experiment), we place the current region into an updating pool. After examining all the decisions, we assess the updating pool to identify pairs of countries with similar populations and capital. If such pairs are found, we swap their groups, remove them from the updating pool, and reset their number of inconsistencies.

\section{Evaluation Results}

In this section, we present the results of a series of experiments conducted to validate the effectiveness of our proposed dynamic grouping negotiation strategy. Main results can be found in Table 2\footnote{Due to the page limitation, we provide additional evaluation results on our GitHub repository (https://github.com/GeminiLn/AI4GCC.git) and Leaderboard}. Specifically, we compare our results with the baseline method and several ablation strategies derived from our proposed method. The methods we employed are listed below:

\textbf{Baseline model (w/o negotiation)}: In this baseline model, negotiation is turned off. Under this setting, there is no negotiation across regions, and each region focuses solely on its own interests.

\textbf{Bilateral negotiation}: The bilateral negotiation method follows the default setting from the official GitHub repository.

\textbf{Static grouping negotiation (mitigation-only/mitigation and saving)}: The static grouping negotiation consists of 9 groups that remain static throughout the entire training process. To further investigate the value of economic constraints like saving, we test both the mitigation-only method and the method that negotiates and evaluates mitigation and saving simultaneously (as described earlier).

\textbf{Dynamic grouping negotiation (mitigation and saving)}: The dynamic grouping negotiation method with mitigation and saving adheres to the entire negotiation protocol discussed in the methodology development section.

\begin{table}[h]
\centering
\begin{tabular}{lccccccc}
\toprule
Method & Temp. Rise & Gross Output & Climate Index & Econ. Index & Hypervolume Contribution  \\
\midrule
Baseline model & 4.47 & 6608 & 0.44 & 0.66 & 0.29  \\
Bilateral negotiation & 1.93 & 4333 (11712) & 0.88 & 0.41 (1.22) & 0.36 (1.08) \\
Static (Mitig.) & 3.89 & 5922 (12092) & 0.54 & 0.59 (1.27) & 0.32 (0.93) \\
Static (Mitig.\&Saving) & 3.41 & 7569 (21734) & 0.62 & 0.77 (2.33) & 0.48 (1.44) \\
Dynamic & 3.79 & 7630 (29125) & 0.56 & 0.77 (3.15) & 0.43 (1.76) \\

\bottomrule
\end{tabular}
\caption{Evaluation Results (The results before April 3. modification are in parentheses)}
\label{tab:methods-results}
\end{table}

As illustrated in Table 2, our group negotiation strategy surpasses other approaches in terms of gross output, achieving values of 7569 for static grouping and 7630 for dynamic grouping, both significantly higher than any other method. Furthermore, the static grouping strategy, which takes into account both mitigation and saving rate, attains the highest hypervolume contribution (0.48). It is crucial to mention that the dynamic grouping strategy does not outperform the static grouping strategy. This occurs after the evaluation method's revision on April 3, which, in fact, weakens the impact of the economic index. For reference, we retain the original evaluation results in parentheses, as they serve as our primary optimization direction during the competition. It is also worth noting that our method's results exhibit slightly higher temperature rises (3.41 and 3.79) compared to some other approaches, such as bilateral negotiation (1.87) and static mitigation (2.81). These results reveal a trade-off between temperature rise and gross output, making it difficult to determine whether to prioritize reducing temperature rise or increasing economic gains. During our evaluation, we observed that the mitigation rate in the bilateral negotiation setting yielded a uniform and unusually high value (approximately 0.9 for each region), which contributed to the low-temperature rise. We believe it is unrealistic in real-world scenarios to expect each region to maintain a 90\% mitigation rate. Consequently, we aimed to maintain a realistic mitigation rate while promoting regional development under the premise of protecting the global environment. In summary, although we lack sufficient time to optimize the dynamic grouping strategy based on the new hypervolume, our methods continue to demonstrate superior overall performance across the evaluation criteria while maintaining feasibility in real-world contexts.

\section{Conclusion and Future Work}

Our proposal aligns with the well-being principle and responsibility principle in the Montreal Responsible AI Declaration. Broadly, fighting climate change aims to enhance the living conditions of people from all over the world and enforce government accountability. In particular, our approach adheres to the democratic participation principle in that we believe that the unit in climate negotiations should not be confined to individual countries, as regional climate and economic requirements may differ. To achieve this, we proposed establishing regional partnerships to allow countries to democratically deliberate on how to adjust their carbon emissions while considering the complexities of the economic landscape. Our assumption is that less powerful countries can form groups to increase their voice when negotiation necessitates economic deals as well. Countries can discuss in a regional unit for climate deals that would better fit their needs economically, meanwhile reducing the carbon footprints.

We hope that our work would improve the \texttt{RICE-N} model and inspire future model development to simulate global climate change negotiation that emphasizes the grouping among regional grouping of similar countries. As we have demonstrated, such consideration not only reduces temperature rise but also improves economic gains among countries, making it a practical measure for adoption in the future. Moving forward, we will continue refining our negotiation strategy, including integrating the dynamic grouping approach into the negotiation process, finding more effective ways to balance temperature rise and economic gains, and exploring additional potential negotiation strategies.

\bibliography{neurips_2023}
\bibliographystyle{plainnat}

\end{document}